\documentclass[preprint,12pt]{elsarticle}

\usepackage{amssymb}
\usepackage{amsmath}
\usepackage{textcomp}
\usepackage{stfloats}
\usepackage{url}
\usepackage{verbatim}
\usepackage{cite}
\usepackage{amsmath,amssymb,amsfonts}
\usepackage{graphicx}
\usepackage{textcomp}
\usepackage{subfig}
\usepackage{dsfont}
\usepackage{color, soul}
\usepackage{hyperref}
\usepackage{multirow}
\usepackage{float}
\usepackage{pifont}  
\usepackage{booktabs}
\usepackage{float}
\usepackage{blindtext}
\journal{Expert Systems with Applications}

\begin{document}

\begin{frontmatter}




\title{Unsupervised Deformable Image Registration with Local-Global Attention and Image Decomposition}
\author[a,b]{Zhengyong Huang} 
\author[c]{Xingwen Sun} 
\author[d]{Xuting Chang} 
\author[a,b]{Ning Jiang} 
\author[e]{Yao Wang} 
\author[f]{Jianfei Sun} 
\author[a,c]{Hongbin Han} 
\author[a,b,g]{Yao Sui\corref{cor1}} 
\ead{yaosui@pku.edu.cn}
\cortext[cor1]{Corresponding author.}
\affiliation[a]{organization={Institute of Medical Technology, Peking University Health Science Center},
            city={Beijing},
            country={China}}
\affiliation[b]{organization={National Institute of Health Data Science, Peking University},
            city={Beijing},
            country={China}}
\affiliation[c]{organization={Department of Radiology, Peking University Third Hospital},
            city={Beijing},
            country={China}}
\affiliation[d]{organization={Department of Pediatrics, Peking University First Hospital},
            city={Beijing},
            country={China}}
\affiliation[e]{organization={Pediatric Epilepsy Center, Peking University First Hospital},
            city={Beijing},
            country={China}}
\affiliation[f]{organization={School of Biological Science and Medical Engineering, Southeast University},
            city={Nanjing},
            country={China}}
\affiliation[g]{organization={Institute for Artificial Intelligence, Peking University},
            city={Beijing},
            country={China}}

\begin{abstract}
Deformable image registration is a critical technology in medical image analysis, with broad applications in clinical practice such as disease diagnosis, multi-modal fusion, and surgical navigation. Traditional methods often rely on iterative optimization, which is computationally intensive and lacks generalizability. Recent advances in deep learning have introduced attention-based mechanisms that improve feature alignment, yet accurately registering regions with high anatomical variability remains challenging.
In this study, we proposed a novel unsupervised deformable image registration framework, LGANet++, which employs a novel local-global attention mechanism integrated with a unique technique for feature interaction and fusion to enhance registration accuracy, robustness, and generalizability. We evaluated our approach using five publicly available datasets, representing three distinct registration scenarios: cross-patient, cross-time, and cross-modal CT-MR registration.
The results demonstrated that our approach consistently outperforms several state-of-the-art registration methods, improving registration accuracy by 1.39\% in cross-patient registration, 0.71\% in cross-time registration, and 6.12\% in cross-modal CT–MR registration tasks.
These results underscore the potential of LGANet++ to support clinical workflows requiring reliable and efficient image registration.
The source code is available at \url{https://github.com/huangzyong/LGANet-Registration}.
\end{abstract}


\end{frontmatter}



\section{Introduction}
\label{sec: introduction}
Deformable image registration is critically important in a wide range of medical image processing and analysis tasks \citep{chen2019real,zheng2024deep}, such as motion correction \citep{sui2025unsupervised}, multi-modal image fusion \citep{yang2025rethinking}, disease diagnosis \citep{lv2022joint}, and image-guided surgery \citep{li2023robust}. The goal of image registration is to determine a transformation that spatially aligns an image (moving image) with its reference image (fixed image). The accuracy of this process is paramount, as it directly impacts the reliability of downstream analyses. Unfortunately, developing a registration method that consistently delivers highly accurate alignments in various scenarios remains challenging \citep{song2022cross}.

Deformable image registration is typically approached using iterative optimization techniques that focus on the repetitive enhancement of an empirically formulated energy function based on warped and fixed images \citep{meng2023non}. However, the inherently time-consuming nature of this iterative methodology poses a significant challenge to its application in real-time clinical settings, such as image-guided intraoperative navigation \citep{liebmann2024automatic}.

Recently, deep learning methodologies have emerged as exceptionally rapid solutions for deformable image registration \citep{liu2021advances,zhong2023unsupervised}, exhibiting notable efficacy due to their advanced feature learning capabilities in contrast to traditional techniques. Unsupervised learning paradigms, which leverage similarity metrics without requiring ground-truth deformation fields, have garnered significant interest for their practicality and scalability \citep{balakrishnan2019voxelmorph}. These approaches \citep{meng2023non} generally begin by developing a parameterized mapping function from a pair of fixed-moving images to the associated deformation field. The parameters are then refined using a dataset comprising unregistered pairs. Upon completion of the training phase, a deformation field can be quickly inferred through a single feed-forward pass. However, in instances where the displacement between image pairs is substantial, estimating the deformation field becomes challenging and often renders these learning-based methods ineffective \citep{zhao2019unsupervised}. Several studies have identified this issue, highlighting that the efficacy of direct estimation methods is generally limited in demanding clinical scenarios involving large displacements \citep{kang2022dual, lewis2020fast}.

To address this issue, researchers adopted a coarse-to-fine learning approach, which decomposes the target deformation field into a series of more readily estimable components \citep{li2024coarse}. This decomposition process facilitates several successive warps of the moving image along with error corrections. As a result, misalignments in the deformation fields from the prior decomposition can be corrected in subsequent estimations, transitioning from coarse to fine accuracy. Depending on the configurations of the implemented models, these techniques typically fall into two primary categories: iterative registration and pyramid registration. Iterative techniques \citep{zhao2019recursive,hu2022recursive} recurrently refine the deformation field but suffer from high computational costs due to repeated feature extraction at each step. Pyramid-based methods \citep{lv2022joint,meng2023non}, which construct feature pyramids at different resolutions, offer a more efficient alternative by first estimating a coarse field at low resolution and progressively refining it.

Simultaneously, attention mechanisms have garnered substantial interest within the medical image registration community due to their straightforward yet effective designs \citep{leroy2023structuregnet,kong2023indescribable}. Extensive studies indicate that, in comparison to other techniques \citep{liu2022coordinate,ruhaak2017estimation}, the utilization of attention-based techniques has enhanced registration performance.
For instance, TransMorph \citep{chen2023transmatch} employs transformers for global feature correlation, while GroupMorph \citep{tan2024groupmorph} utilizes group-wise correlation to capture large deformations and small deformations. However, a limitation of these methods is that the interaction between features from the moving and fixed images is often insufficiently explored, which hinders the model's ability to learn precise, voxel-level correspondences.

Building upon these successes, we propose LGANet++, a novel approach based on a pyramid registration framework. We introduce a local-global attention module (LGAM) that captures both fine-grained local correspondences and long-range contextual relationships to handle significant regional variations in deformation. Furthermore, we design a feature interaction and fusion module (FIFM) to enhance information exchange between the warped and fixed images, along with a multi-scale fusion module (MSFM) to integrate semantic cues across different resolutions. These components collectively enable hierarchical refinement of the deformation field, significantly improving the robustness of registration across diverse tasks. 
Extensive experiments demonstrate that our method consistently achieves high accuracy and strong generalization in various registration tasks. Quantitative results show that LGANet++ outperforms state-of-the-art methods, with registration accuracy improved by 1.39\% in cross-patient registration, 0.71\% in cross-time registration, and 6.12\% in cross-modal CT–MR registration tasks.
Notably, these improvements highlight the superior effectiveness and robustness of LGANet++ in challenging registration scenarios, particularly for cross-modal CT-MR registration where large appearance discrepancies exist.

Our primary contributions are detailed below:
\begin{itemize}
\item[$\bullet$] We propose a novel coarse-to-fine encoder-decoder network, LGANet++, incorporating a local-global attention mechanism for accurate and robust deformable image registration.
\item[$\bullet$] We design a multi-scale fusion module (MSFM) that effectively integrates and transfers semantic information across feature maps of different resolutions, enhancing contextual coherence in the deformation field.
\item[$\bullet$] We develop two dedicated modules: LGAM for capturing both local and global feature dependencies, and a feature interaction and fusion module (FIFM) comprising an image decomposition module (IDM) and a channel-wise attention module (CWAM) for structured and refined alignment.
\item[$\bullet$] We conduct extensive experiments on five datasets across three distinct scenarios to assess the efficacy of our proposed method, showcasing its superior registration performance compared to nine state-of-the-art techniques.
\end{itemize}

Our preliminary results have been reported in \citep{huang2024dual}, while the current study incorporates a substantial extension in both the methodology and the scope of the experiments. The remainder of this paper is organized as follows. Section \ref{methods} details the architecture of the proposed LGANet++ framework and its core components. Section \ref{experiments} describes the experimental setup, datasets, and presents a comprehensive comparison with state-of-the-art methods, followed by ablation studies. Section \ref{discussion} discusses the strengths, limitations, and conclusions of our work, along with suggestions for future research.

\begin{figure*}[t]
    \centering
    \includegraphics[width=\linewidth]{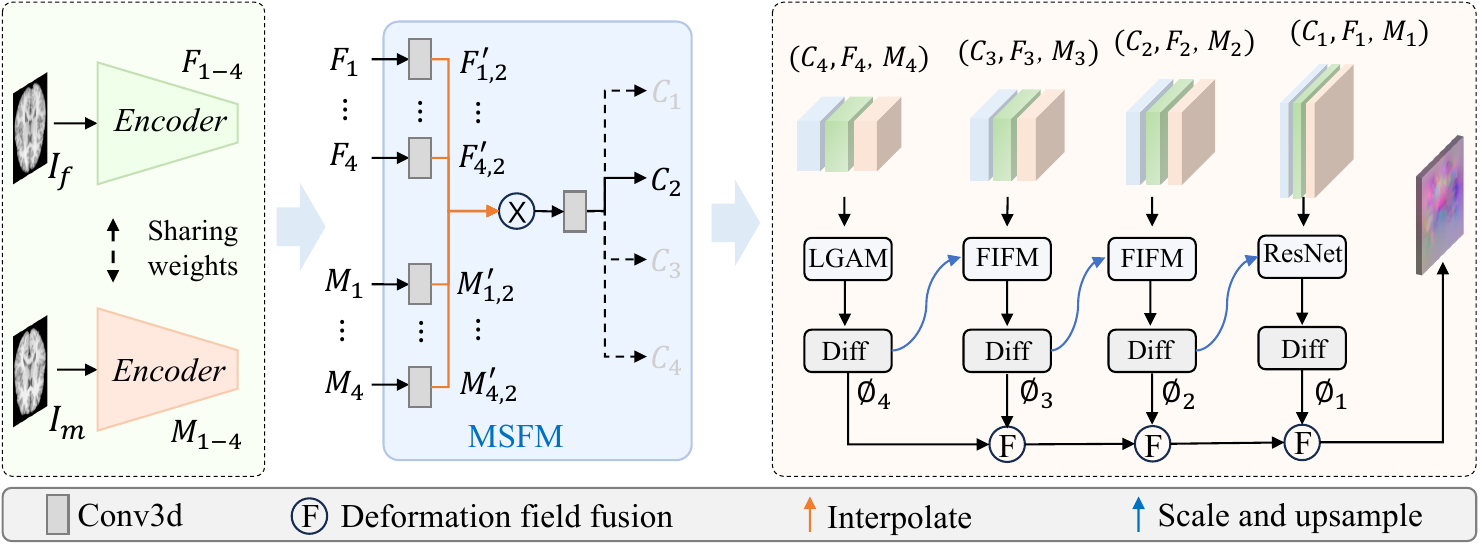}
    \caption{Illustration of our proposed LGANet++. Two structurally identical encoders (sharing weights) extract features from the moving image and the reference image, respectively. A multi-scale fusion module then combines these two feature sets. Finally, the decoder performs deformation field ($\phi_i$) estimation based on the fused features ($C_i$), moving image features ($M_i$), and reference image features ($F_i$). The deformation fields are progressively optimized from $\phi_4$ to $\phi_1$ in a coarse-to-fine manner, with $\phi_1$ representing the final deformation field. MSFM: multi-scale fusion module, which is used to fuse moving image features and fixed image features. LGAM: local-global attention module, which is used to estimate the initial deformation field $\phi_4$. FIFM: feature interaction and fusion module, which is used to estimate and refine the deformation field at each stage of the decoding process. Diff: diffeomorphic layer, which ensures smooth, invertible, and topology-preserving transformations by integrating a differentiable exponential mapping into the deformation field prediction.}
    \label{fig: network}
\end{figure*}

\section{Methods}\label{methods}
Deformable image registration seeks to identify a deformation field $\phi$ that minimizes the disparity between a moving image $I_m$ and a fixed image $I_f$ \citep{liu2023co}. In this study, we adopt an unsupervised learning approach, where the deformation field is predicted directly from the input image pair through a parametric network $\mathcal{F}_\theta$:
\begin{equation}
    \phi = \mathcal{F}_{\theta}(I_f, I_m),
\end{equation}
and the warped image $I_w$ is then obtained using a spatial transformer networks (STN) \citep{jia2023fourier} as $I_w = I_m \circ \phi$.
In this context, $I_m, I_f, I_w \in \mathcal{R}^{H \times W \times D}$, and $\phi \in \mathcal{R}^{3 \times H \times W \times D}$ denotes the displacement-based deformation field. $\circ$ denotes the warp operation, which involves resampling and interpolating moving images using the STN module. \citep{balakrishnan2019voxelmorph}.

Our proposed framework, LGANet++, is designed to estimate $\phi$ in a coarse-to-fine manner using a pyramid registration strategy.
As depicted in Fig. \ref{fig: network}, LGANet++ consists of three principal components: (1) a dual-stream encoder for multi-scale feature extraction, (2) a multi-scale fusion module for integrating pyramidal feature maps and transferring semantic information across different resolutions, and (3) a decoder that utilizes local-global attention mechanisms in conjunction with feature interaction and fusion for hierarchical deformation refinement.
Detailed descriptions of each module are provided in the following subsections.

\subsection{Dual-stream Feature Encoder}
To achieve pyramid registration, we implement a dual-stream feature encoder with shared weights to derive pyramidal feature maps from both the fixed image $I_f$ and the moving image $I_m$, in a manner analogous to the approach outlined in \citep{wang2024recursive}. This encoder comprises four convolutional blocks, starting with eight kernels in the initial block. 
As shown in Fig. \ref{fig: network}, the feature maps produced by the encoder are labeled as $F_i$ and $M_i$ for $i\in [1, 2, 3, 4]$. Moving from $F_1(M_1)$ to $F_4(M_4)$, each step results in a halving of both the resolution and the size of the feature maps, while the number of feature channels simultaneously doubles.

\subsection{Multi-scale Fusion Module}
To effectively combine information across scales, we introduce a multi-scale fusion module.
First, we apply a convolutional layer to reduce the channels of $F_i(M_i)$ to $c$, where $c$ is set to eight. Then, we adjust the size of the feature maps at different levels to match the target size $s_t$:
\begin{equation}
F_{k,i}' = 
    \begin{cases} 
    \mathcal{F}_{inter}(\mathcal{F}_{conv}(F_i)) & \text{if } s_i < s_t, \\ 
    \mathcal{F}_{conv}(F_i) & \text{if } s_i = s_t, \\ 
    \mathcal{F}_{pool}(\mathcal{F}_{conv}(F_i)) & \text{if } s_i > s_t, 
    \end{cases}
\end{equation}
and similarly for $M_{k,i}^\prime$ .And the fused feature map $C_i$ at level $i$ is obtained by multiplying all rescaled feature maps followed by a convolutional operation:
\begin{equation}
C_i = \mathcal{F}_{conv}(F'_{1,i} \otimes F'_{2,i} \otimes F'_{3,i} \otimes F'_{4,i} \otimes M'_{1,i} \otimes M'_{2,i} \otimes M'_{3,i} \otimes M'_{4,i}) 
\end{equation}
where $s_i$ denotes the size of $F_i$ or $M_i$. $\mathcal{F}_{conv}$ is convolutional operation. $C_i$ is the same size as $F_i$. $\otimes$ denotes the multiplication operation. $\mathcal{F}_{pool}$ and $\mathcal{F}_{inter}$ denote global average pooling and trilinear interpolation operations, respectively.
$F'_{k,i}$ and $M'_{k,i}$ denote the rescaled results of the feature maps from the $k$-th level according to the size requirement $s_t$ of the $i$-th level.

\subsection{Local-Global Attention Module}
Following the encoding process, a decoder is employed to integrate image features and learn the correspondences necessary to estimate the deformation field at various scales. 
A key component within the decoder is the Local and Global Attention Module (LGAM), which is designed to capture both fine local details and global contextual relationships.
The LGAM consists of a local attention mechanism that computes self-attention independently within different regions of the image. This design effectively handles the substantial heterogeneity in deformations across local areas of the aligned image. Additionally, a global attention mechanism is incorporated to maintain coherence and interactions between these regions.
Ultimately, the LGAM is proposed to establish a reliable initial deformation field $\phi_4$. 

\begin{figure}[t]
    \centering
    \includegraphics[width=\linewidth]{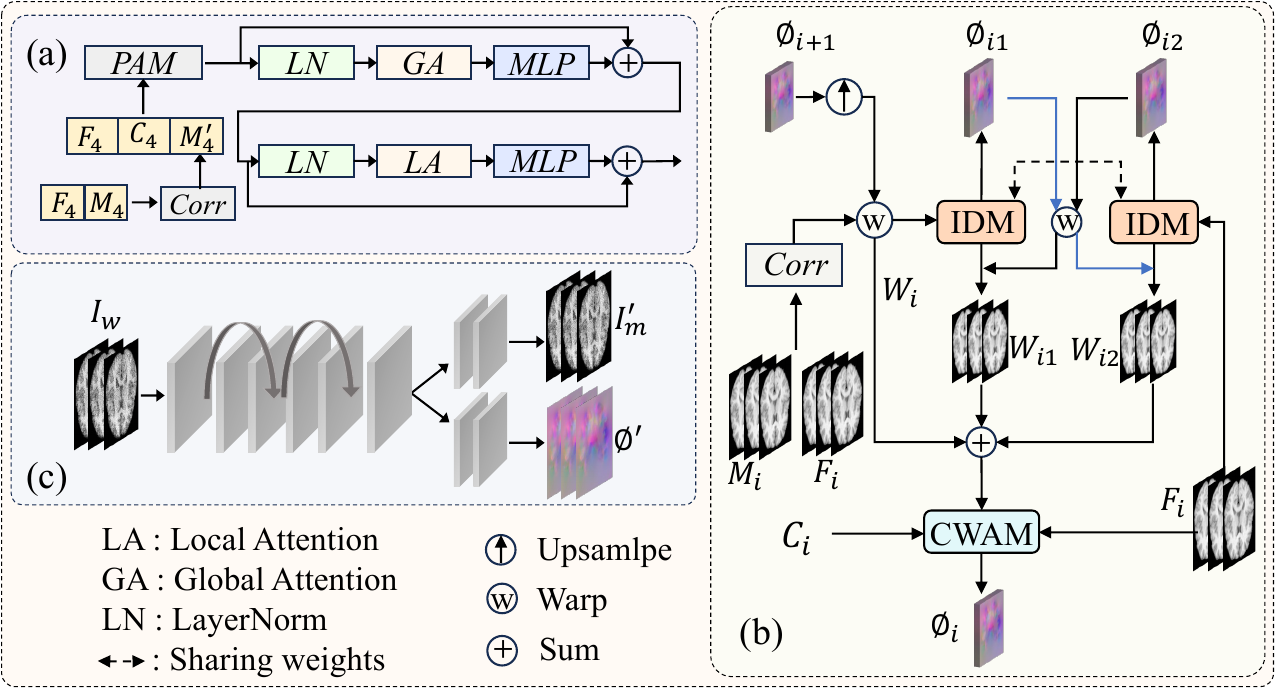}
    \caption{Depiction of the proposed local and global attention module (LGAM) and the feature interaction and fusion module (FIFM). (a) Computational procedure of LGAM. Features from different branches are concatenated and then processed through a positional attention module (PAM) to capture spatial dependencies. This is followed by attention mechanisms operating at both global and local levels to extract integrated and fine-grained features. (b) Feature interaction and fusion module (FIFM). $Corr$ indicates a 3D correlation layer that computes pixel-wise similarity relationships between the two feature maps. (c) Image decomposition module (IDM).}
    \label{fig: sub-nets}
\end{figure}

As illustrated in Fig. \ref{fig: sub-nets}(a), the LGAM takes $F_4$, $M_4$, and $C_4$ as inputs. First, the 3D correlation layer \citep{meng2024correlation} computes a correlation map $M'_4$ between $F_4$ and $M_4$. These features are concatenated and processed via a position attention module (PAM) followed by layer normalization (LN). 
The configuration of PAM is depicted in Fig. \ref{fig: pam-cwam}(a). Subsequently, we determine the global attention (GA) maps for the features and derive the final global attention output, followed by a multi-layer perceptron (MLP) that comprises two fully connected layers. Furthermore, we incorporate a local attention (LA) module, which separates the feature maps into several volumes and then computes the local attention maps for each volume. This method effectively aids in capturing the matching relations of local features. 
Residual connections are used in both GA and LA to facilitate training. Accordingly, the attention mechanism for LA and GA is defined as follows:
\begin{equation}
\begin{aligned}
    Q = Proj_q(X), \quad K = & Proj_k(X), \quad V = Proj_v(X), \\
    Attention(Q,K,V) = & \text{Softmax}(QK^T / \sqrt{l})V,
\end{aligned}
\end{equation}
where $Proj(\cdot)$ signifies a linear projection and $X$ denotes the input feature maps. Initially, $X \in \mathcal{R}^{H \times W \times D \times C}$ is transformed into feature vectors $X \in \mathcal{R}^{N \times C}$ using $N = H \times W \times D$. For LA, $X$ is restructured to $X \in \mathcal{R}^{m \times n \times C}$, where $m$ indicates the number of local volumes and $n = h \times w \times d$ represents the dimensions of these volumes. The scaling factor $l$ is associated with the dimension of $K$.

\subsection{Feature Interaction and Fusion Module}
\begin{figure}[h]
    \centering
    \includegraphics[width=0.6\linewidth]{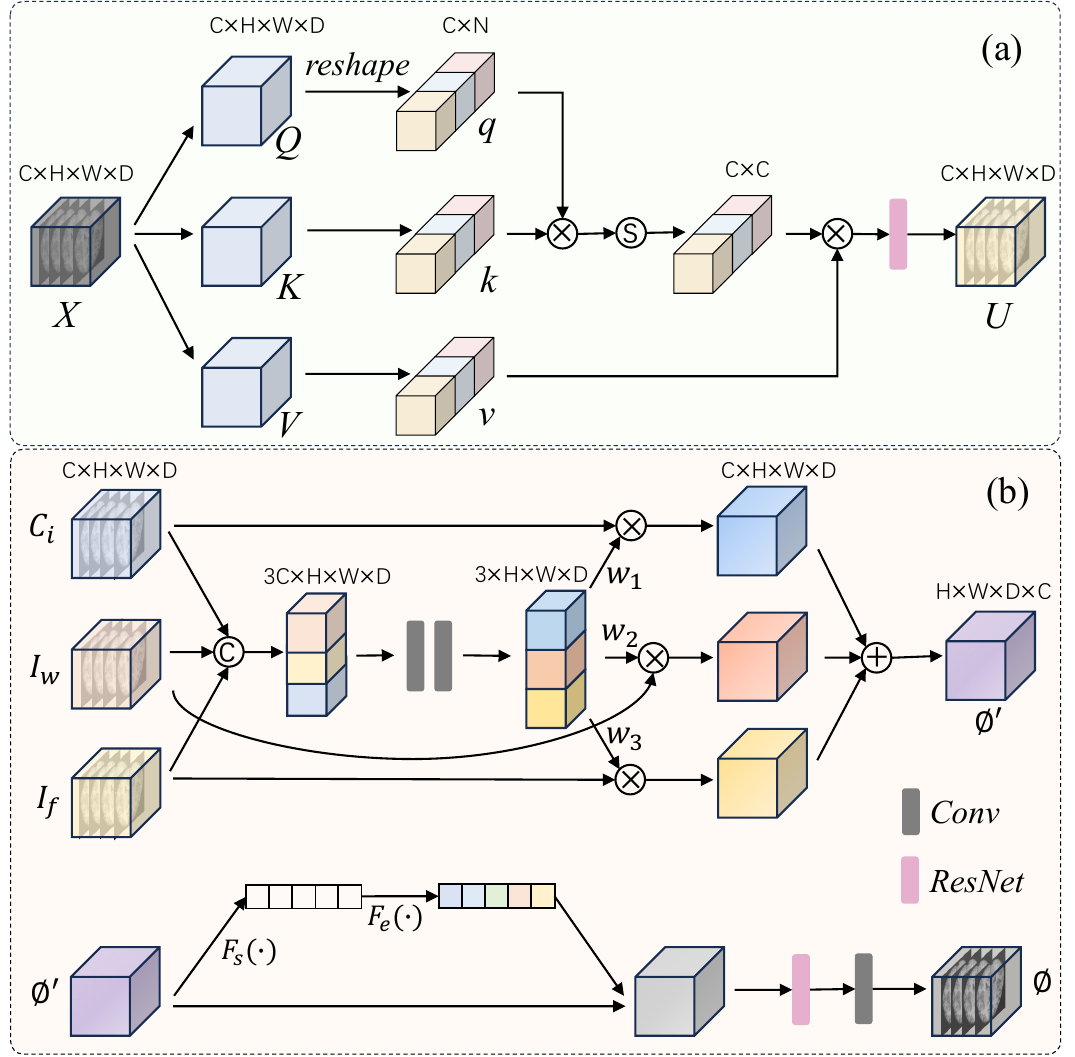}
    \caption{(a) Architecture of the Position Attention Module (PAM). The input features are processed through convolutional layers (blue arrows) to generate Q, K, V representations. The self-attention mechanism is applied after reshaping, followed by a Softmax operation ($\textcircled{s}$) to compute spatial dependencies. The output is then reconstructed and enhanced via a residual block to produce refined features $U$. (b) Structure of the Channel-wise Attention Module (CWAM). The top part illustrates the adaptive channel fusion block, which learns channel-wise weights for feature maps $C_i$, $I_w$, and $I_f$ through a multi-channel attention (MCA) mechanism. The bottom part consists of a SENet-based squeeze-excitation block, a ResNet block, and a convolutional layer, collectively used to selectively emphasize informative channels and output the final deformation field $\phi$.}
    \label{fig: pam-cwam}
\end{figure}

Fig. \ref{fig: network} illustrates the process of obtaining the preliminary deformation field $\phi_4$ via LGAM, which subsequently allows for the generation of the warped image $I_w$ from the moving image $I_m$ ($I_w = I_m \circ \phi$). During each stage of decoding, we develop a feature interaction and fusion module (FIFM) to integrate the feature maps $C_i$, $M_i$, and $F_i$ to progressively refine the quality of the deformation field, transitioning from lower to higher resolution. As depicted in Fig. \ref{fig: sub-nets}(b), the $M_i$ and $F_i$ are fed into a 3D correlation layer to obtain a 3D correlation map $M_i'$. 
Then, the input $\phi_{i+1}$ is upsampled to fit the dimensions of $M_i'$ and subsequently applied to $I_m'$ to produce $W_i$ ($W_i = M_i' \circ (\mathcal{F}_{up}(\phi_{i+1}))$), where $\mathcal{F}_{up}(\cdot)$ defines the upsampling operation. Thereafter, the image decomposition module (IDM) is used to achieve alignment between the warped and fixed images by enforcing consistency in their decoupling results. 
Then, we employ the channel-wise attention module (CWAM, illustrated in Fig. \ref{fig: pam-cwam}(b)) to integrate the features from $C_i$, $F_i$, and the outputs of the IDM. The process is defined as follows:
\begin{equation}
    \begin{aligned}
        \phi_{i1}, W_{1} &= IDM(W_i),
        \phi_{i2}, W_{2} = IDM(F_i), \\
        W_{i1} &= W_{1} \circ \phi_{i2},
        W_{i2} = W_{2} \circ \phi_{i1}, \\
        \phi_i' &= CWCA(F_i, C_i, (W_i+W_{i1}+W_{i2})), \\
        \phi_i &= \phi_i' + \text{STN}(\phi_{i1}, \phi_i') + \text{STN}(\phi_{i2}, \phi_i').
    \end{aligned}
\end{equation}
Moreover, we adopt the approach outlined in \citep{vercauteren2009diffeomorphic, wang2024recursive} by integrating diffeomorphic layers with a transform layer. Diffeomorphic deformations are characterized by their smoothness and reversibility, ensuring the preservation of topological structures. Our specific approach involves defining a minuscule time step of $1 / 2^t$, along with $t = 7$ in our implementation, and applying a recurrent warping technique: 
\begin{equation}
\label{equa: 5}
    \phi^{{t-1}} = \phi^{t} + \phi^{t} \circ \phi^{t},
\end{equation}
where the recurrence is initialized initial with $\phi^{t} = \phi_i / 2^t$. Upon the completion of \textit{t} iterations, and after performing \textit{t} iterations, the final deformation field is obtained as:
\begin{equation}
    \phi_i = \phi^{0}.
\end{equation}

\subsubsection{Position Attention Module}
Fig.~\ref {fig: pam-cwam}(a) illustrates the architecture of PAM, where the input $X$ is derived from the concatenation of $F_4$, $C_4$, and $M_4'$ (as shown in Fig.~\ref{fig: sub-nets}(a)). Initially, feature extraction is performed through three convolutional steps to generate query ($Q$), key ($K$), and value ($V$) ($Q, K, V\in \mathcal{R}^{C \times H \times W \times D}$). Subsequently, $Q, K, V$ are reshaped into $q, k, v \in \mathcal{R}^{C \times N}$, with $N$ representing $H \times W \times D$. A self-attention mechanism is then applied, and its output is transformed back into the original input dimensions $X \in \mathcal{R}^{C \times H \times W \times D}$. Finally, a ResNet block is applied to produce $U \in \mathcal{R}^{C \times H \times W \times D}$. The detailed computation process of PAM is delineated as:
\begin{equation}
\label{eq: pam}
    \begin{aligned}
        Q, K, V &= Conv_{q, k, v}(X), \\
        q, k, v &= \mathcal{F}_r(Q, K, V), \\
        U = Conv_{res}(\mathcal{F}_r(&\text{Softmax}(q \cdot k^T / \sqrt{l}) \cdot v)),
    \end{aligned}
\end{equation}
$Conv_{q,k,v}$ signifies a standard convolution process, while $Conv_{res}$ represents a ResNet block. The operation $\mathcal{F}_r(\cdot)$ denotes a reshape transformation. The scaling factor $l$ is associated with the dimension of $k$.

\subsubsection{Channel-wise Attention Module}
The CWAM configuration is illustrated in Fig. \ref{fig: pam-cwam}(b), which consists of two primary components: the multi-channel attention (MCA) block \citep{wang2021head}, located at the top of Fig. \ref{fig: pam-cwam}(b), and the squeeze-expansion (SE) block \citep{hu2018squeeze}, positioned at the bottom of Fig. \ref{fig: pam-cwam}(b). The MCA fuses features adaptively by emphasizing important channels across $C_i$, $I_w$, and $I_f$. The intermediate output $W \in \mathcal{R}^{3 \times H \times W \times D}$ represents the importance of $C_i, I_w$, and $I_f$. Dividing $W$ into separate channels, we obtain $w_1, w_2, w_3$, which are then multiplied with the respective input features $C_i, I_w, I_f$, before being concatenated to form the output $\phi' \in \mathcal{R}^{3C \times H \times W \times D}$. Following additional feature selection through SENet and ResNet, the deformation field $\phi \in \mathcal{R}^{3 \times H \times W \times D}$ is ultimately derived by reducing feature channels via a convolutional layer.

\begin{figure}[tp]
    \centering
    \includegraphics[width=0.6\linewidth]{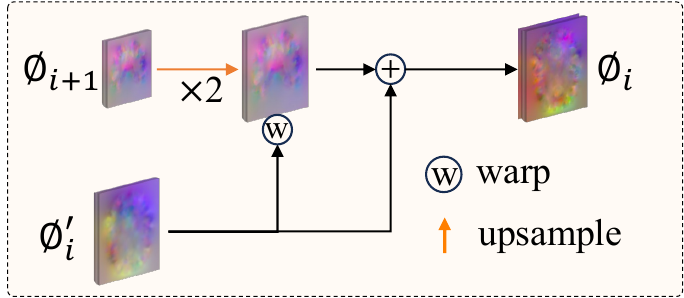}
    \caption{Illustration of the optimization strategy for refining the deformation field in a coarse-to-fine manner. Here, $\phi_{i}'$ denotes the initial output deformation field in the current stage. $\phi_{i}$ signifies the deformation field acquired subsequent to the optimization process using the previous stage $\phi_{i+1}$, which possesses a lower resolution and captures relatively coarser deformation characteristics.}
    \label{fig: field-refine}
\end{figure}

\subsection{Coarse to Fine Optimization}
This section outlines the optimization procedure for the deformation field that transitions from coarse to fine resolution. Our proposed LGANet++ constructs a succession of four consecutive deformation fields with progressively finer resolutions, denoted as $[\phi_4, \phi_3, \phi_2, \phi_1]$. As illustrated in Fig. \ref{fig: network}, $\phi_4$ emerges as the output of LGAM, serving as an initial deformation field that primarily captures the coarse deformation characteristics. 
The deformation field at each subsequent stage is refined by incorporating an up-sampled version of the previous stage’s deformation field, scaled by a factor of 2. This up-sampled field is then warped by the deformation field currently being estimated, followed by a summation process. These up-sampling and warping steps are iteratively and sequentially applied to produce the final deformation field, as depicted in Fig. \ref{fig: field-refine}, which captures extensive multi-scale information. This approach enables the model to propagate robust contextual information across hierarchical decoding layers, where the estimated deformation fields are progressively refined from coarse to fine resolution, thereby integrating both high-level contextual details and low-level specific features. 

The optimization procedure is characterized by the expression:
\begin{equation}
\label{equa: 9}
    \phi_{i} = \mathcal{F}_{up}(\alpha \times \phi_{i+1}) \circ \phi_{i}' + \phi_{i}',
\end{equation}
where $\mathcal{F}_{up}(\cdot)$ represents the up-sampling operation. The scaling factor $\alpha$ for up-sampling is assigned a value of 2.

\subsection{Loss function}
To facilitate network training, we employ local normalized cross-correlation (NCC) with a specified window size of $\omega$ ($\omega$=9 in this study) to assess the discrepancy between the warped image $I_w$ and the fixed image $I_f$. The range of NCC values extends from -1 to 1. For a given local window, NCC is calculated as:
\begin{equation}
    NCC(I_f, I_w) = \sum \frac{(I_f - \mu_{f})(I_w - \mu_{w})}{\sigma_{f} \sigma_{w}},
\end{equation}
where $\mu_{f}$ and $\mu_{w}$ are the mean values within the local window for the fixed and warped images, respectively, while $\sigma_{f}$ and $\sigma_{w}$ represent the standard deviations.

A high NCC is essential for accurate registration. Thus, the loss function is formulated as:
\begin{equation}
    \mathcal{L}_{sim}(I_f, I_w) = 1 - NCC(I_f, I_w).
\end{equation}
To maintain the spatial continuity of the resulting deformation field, a regularization term founded on the gradient of the deformation field is incorporated. The comprehensive loss function is expressed as:
\begin{equation}
    \mathcal{L}(I_f, I_m, \phi) = \mathcal{L}_{sim}(I_f, I_m \circ \phi) + \lambda ||\nabla \phi||^2,
\end{equation}
where the weight factor $\lambda$ is set to 1 to ensure fair comparison with previous works \citep{balakrishnan2019voxelmorph}.

\section{Experiments and Results}\label{experiments}
\subsection{Experimental Setup}
\subsubsection{Datasets and Pre-processing}
Our proposed registration approach, LGANet++, was evaluated using five publicly available datasets: LPBA \citep{shattuck2008construction}, IXI \citep{wang2024recursive}, OASIS \citep{marcus2007open}, Lung CT \citep{hering2020learn2reg}, and Abdomen CT-MR \citep{linehan2016cancer,xu2016evaluation, CHAOSdata2019}.

\emph{\textbf{LPBA Dataset}}
The LONI Probabilistic Brain Atlas (LPBA) comprises T1-weighted MRI scans of 40 human subjects, with 56 anatomically complex brain regions manually delineated by trained raters following the standardized LPBA protocol. For all datasets, skull removal was executed using FreeSurfer. Each volume was subject to center cropping to dimensions of $144 \times 176 \times 144$ $(1mm\times1mm\times1mm)$. The dataset was then split into sets consisting of 30 for training purposes and 10 for testing.

\emph{\textbf{IXI Dataset}} The Information eXtraction from Images (IXI) comprises 576 MRI scans. We divided 403, 58, and 115 into training, validation, and test sets. Each scan was resampled to $96\times112\times80$ $(2mm\times2mm\times2mm)$ and segmented into 30 cortical ROIs.

\emph{\textbf{OASIS Dataset}} 
The Open Access Series of Imaging Studies (OASIS) includes a total of 414 T1-weighted brain MRI volumes, each segmented into 35 cortical regions. Each volume was resampled to a dimension of $96 \times 112 \times 80$ $(2mm\times2mm\times2mm)$. In this study, the OASIS was used for an external test.

\emph{\textbf{Lung CT Dataset}}
This dataset serves to evaluate our proposed approach for cross-time registration tasks. It comprises 30 pairs of lung CT images, featuring scans from 30 participants taken during both exhalation and inhalation, resulting in one exhalant volume and one inhalant volume per pair. Each scan was center-cropped to dimensions of $96 \times 192 \times 192$. For model evaluation, we split the dataset into two subsets: 20 scans for training and 10 for testing. 

\emph{\textbf{Abdomen CT-MR Dataset}}
The Abdomen CT-MR dataset comprises 8 pairs of CT-MR images, with 2 allocated for training and 6 reserved for testing. Each image contains four structures annotated manually. In addition, 89 unpaired CT-MR scans are available (consisting of 40 MR scans and 49 CT scans), which serve as supplementary training data. A random selection process created 90 pairs of CT-MR images for network training. All images were cropped to dimensions of $144 \times 160 \times 192$. During preprocessing, the CT images initially underwent clipping at pixel values within the range of $[-800, 800]$. Subsequently, both CT and MR images were normalized using min-max normalization to scale the voxel intensities of each scan between [0, 1]. 

Following the standard protocol for atlas-based registration, we adopt an atlas-to-patient registration setting for the LPBA, IXI, and OASIS datasets.  In this configuration, one representative volume is selected as the fixed atlas, and the remaining subjects are treated as moving images to be aligned to this common coordinate space. 

\subsubsection{Implementation Details}
Our LGANet++ was implemented utilizing the PyTorch framework and deployed on an NVIDIA 4090 GPU, equipped with 24 GB of memory. The Adam optimizer was employed with an initial learning rate of 1e-4 and momentum coefficients were configured as specified in $[0.99, 0.999]$ \citep{wang2024recursive}. A batch size of 1 was used, except for the IXI dataset, which utilized a batch size of 4. The network underwent training for 300 epochs. To stabilize convergence, the learning rate was decayed linearly from its initial value to 1e-6 starting from the $200^{th}$ epoch. The model selected for evaluation was the one derived from the final training epoch.

\subsubsection{Evaluation Metrics}
In the absence of ground-truth deformation fields in unsupervised registration, we evaluate performance indirectly by assessing the anatomical consistency between warped and fixed images. A comprehensive set of widely used metrics is employed to evaluate performance, encompassing both the accuracy of the warped images and the reliability of the deformation field.

Specifically, we reported the Dice Similarity Coefficient (DSC), 95\% Hausdorff Distance (HD95, in mm), Target Registration Error (TRE, in mm), Recall, and Precision to evaluate registration accuracy, while the percentage of voxels with a Negative Jacobian Determinant (NJD) is used to assess the reliability of the deformation field \citep{fitzpatrick2001distribution,guerra2025deep,de2024optimizing,meng2023non}.

The Dice Similarity Coefficient measures the volumetric overlap between the warped and fixed segmentations of the corresponding anatomical structures. It ranges from 0 to 1, with higher values indicating better alignment. It is defined as:
\begin{equation}
\text{DSC} = 2 \times \frac{|X \cap Y|}{|X| + |Y|},
\end{equation}
where \(X\) and \(Y\) denote the set of voxels in the warped and fixed image segmentations of the corresponding anatomical structures, respectively.

The Hausdorff distance (HD) is a metric for quantifying shape similarity, defined as the maximum of the shortest distances between points in two sets. It emphasizes the congruence of segmentation boundaries. To reduce sensitivity to outliers, the 95th percentile Hausdorff Distance (HD95) is often used. This variant calculates the 95th percentile of the surface distances between two segmented volumes, rather than taking the absolute maximum. As a result, it offers a more robust assessment of segmentation boundary discrepancies. It is computed as follows:
\setlength{\lineskip}{0.1pt}
\setlength{\lineskiplimit}{0.1pt}
\begin{equation}
    \begin{split}
        HD(X,Y)=max(d(X,Y),d(Y,X))\\
        d(X,Y)=\max_{x \in X} \left\{\min_{y\in Y}||x-y||\right\}\\
        d(Y,X)=\max_{y \in Y} \left\{\min_{x\in X}||y-x||\right\}
    \end{split}
\end{equation}
where \(d(X, Y)\) is the distance norm of set $X$ and set $Y$, and $||\cdot||$ is the Euclidean distance between points \(x\) and \(y\).

Target Registration Error quantifies the Euclidean distance between corresponding anatomical landmarks in the fixed and warped images. It is defined as follows:
\begin{equation}
\text{TRE} = \sqrt{(x_f - x_w)^2 + (y_f - y_w)^2 + (z_f - z_w)^2},
\end{equation}
where \((x_f, y_f, z_f)\) and \((x_w, y_w, z_w)\) denote landmark coordinates in the fixed and warped images, respectively.

Precision, also known as the positive predictive value, measures the reliability of the positive predictions. It is defined as the ratio of correctly predicted positive observations to the total number of instances predicted as positive. A high precision value indicates a low rate of false positive errors. This metric is calculated as:
\begin{equation}
\text{Precision} = \frac{TP}{TP + FP},
\end{equation}
where \(TP\), \(TN\), \(FP\), and \(FN\) represent true positives, true negatives, false positives, and false negatives, respectively.

Recall quantifies the model's ability to correctly identify all actual positive instances. It is calculated as the proportion of true positives to the sum of true positives and false negatives. This metric is particularly crucial in applications where missing a positive case (i.e., a \textit{FN}) carries a significant cost. It is defined as follows:
\begin{equation}
\text{Recall} = \frac{TP}{TP + FN}.
\end{equation}

Finally, the NJD indicates the proportion of voxels where the deformation field is non-diffeomorphic (i.e., folding). A lower value indicates a smoother and more reliable deformation:

\begin{equation}
\text{NJD} = \frac{1}{N} \sum \mathbb{I}(J_\phi(\mathbf{x}) < 0),
\end{equation}
where \(J_\phi(\mathbf{x})\) is the Jacobian determinant of the deformation field \(\phi\) at voxel \(\mathbf{x}\), \(\mathbb{I}\) is the indicator function, and \(N\) is the total number of voxels.

To evaluate the significance of the observed performance improvements, we conducted pairwise statistical comparisons. Specifically, the Wilcoxon signed-rank test was employed to compare the proposed LGANet++ with each state-of-the-art method independently across all test samples. A p-value of less than 0.05 was considered to indicate a statistically significant difference. Notably, statistical testing was not conducted for the Negative Jacobian Determinant (NJD), as it serves primarily as a secondary indicator of the deformation field's topological plausibility rather than a primary measure of registration accuracy.

\subsubsection{Comparison methods}
We performed comparative evaluations of our proposed technique against a variety of state-of-the-art registration methods, including VoxelMorph \citep{balakrishnan2019voxelmorph}, PRNet++ \citep{kang2022dual}, PCNet \citep{lv2022joint}, PIViT \citep{ma2023pivit}, ModeT \citep{wang2023modet}, CorrMLP \citep{meng2024correlation}, GroupMorph \citep{tan2024groupmorph}, RDP \citep{wang2024recursive}, and LGANet\citep{huang2024dual}. VoxelMorph represents a conventional approach, utilizing a single-stream U-Net architecture to directly estimate the deformation field. PRNet++ employs a dual-stream network to generate feature pyramids from two input volumes, suggesting sequential pyramid registration with a 3D correlation layer. PCNet extends PRNet++ by adding a deformation field integration module, which utilizes previously predicted coarse fields to warp features, thus progressively estimating finer fields. PIViT introduces a pyramid iterative composite structure to address deformation challenges through low-scale iterative registration with a Swin Transformer-based long-distance correlation decoder. ModeT incorporates a motion decomposition transformer to explicitly model diverse motion modalities, leveraging the transformer's inherent capabilities for estimating deformations. CorrMLP introduces a correlation-aware multi-window MLP block, which captures fine-grained multi-range dependence to perform correlation-aware coarse-to-fine registration. 
GroupMorph adopts group-wise correlation to decompose the deformation field into a series of subfields with different receptive field, which preserves more detailed information to measure the similarity between feature pairs, thereby enhancing its ability of handling complex deformation. 
RDP is a pure convolutional pyramid network, which leverages a step-by-step recursion strategy with the integration of high-level semantics to predict the deformation field from coarse to fine while ensuring the rationality of the deformation field. LGANet is the conference version of our proposed method.
All methods were trained under consistent training conditions.

\begin{table}[htp]
  \centering
  \caption{Results of the ablation study for the LGAM, FIFM, and MSFM modules on the LPBA dataset (mean $\pm$ std). The notation \textbf{Bold} indicates the top result.}
  \label{tab: ablation}
        \begin{tabular}{c|ccccc}
        \hline 
        \specialrule{0em}{0.2pt}{0.2pt}
        \hline 
        Methods & LGAM & FIFM & MSFM & DSC(\%) $\uparrow$ & HD95 $\downarrow$ \\ 
        \hline 
        Model 1 & \ding{55} &\ding{55} &\ding{55}          &70.02$\pm$1.81 &5.61$\pm$0.42  \\
        Model 2 & $\checkmark$ &\ding{55} &\ding{55}       &71.19$\pm$1.53 &5.22$\pm$0.38  \\
        Model 3 & $\checkmark$ &$\checkmark$ &\ding{55}    &72.57$\pm$1.09 &5.16$\pm$0.36  \\ 
        Model 4 & $\checkmark$ &$\checkmark$ &$\checkmark$ &\textbf{73.52$\pm$0.09} &\textbf{5.10$\pm$0.28} 
        \\
        \hline 
        \specialrule{0em}{0.2pt}{0.2pt}
        \hline
      \end{tabular}
\end{table}

\subsection{Experimental Results}
This section presents a comprehensive evaluation of our LGANet++ framework across multiple registration scenarios. We begin with an ablation study to validate the contribution of each component, followed by comparisons with state-of-the-art methods on cross-patient, cross-modal, and cross-time registration tasks. To provide deeper insights beyond quantitative metrics, we simultaneously analyze statistical significance and qualitative results, thereby offering a concise yet comprehensive demonstration of our method’s superiority.

\subsubsection{Ablation Study}
We first analyze the contribution of each key component (LGAM, FIFM, and MSFM) in our framework using the LPBA dataset. Specifically, we first construct a basic dual-branch pyramid registration model based on CNN as the baseline (Model 1). Then, we sequentially add the proposed module to verify its effectiveness in enhancing registration performance. Additionally, we have calculated the specific parameter counts for each model in our ablation study: Model 1 (2.936M), Model 2 (10.225M), Model 3 (21.320M), and Model 4 (24.353M). As summarized in Table~\ref{tab: ablation}, the baseline model (Model 1) without these three modules achieves a DSC of 70.02\%. Introducing the Local-Global Attention Module (LGAM) in Model 2 improves the DSC by 1.17\%, indicating its effectiveness in capturing both local and global contextual features essential for handling regional deformations.
The inclusion of the Feature Interaction and Fusion Module (FIFM) in Model 3 further improves the DSC to 72.57\%, underscoring the importance of explicit feature interaction and decomposition learning between the moving and fixed images. Finally, the full model (Model 4) incorporating the Multi-Scale Fusion Module (MSFM) achieves the highest DSC of 73.52\% and the lowest HD95 of 5.10 mm. The MSFM enables effective integration of multi-resolution features, which is particularly beneficial in capturing structures at varying scales. The overall improvement of 3.50\% in DSC and reduction in HD95 demonstrate the complementary nature of these modules.

\begin{table}[h]
    \centering
    \caption{Quantitative results of our approach contrasted with those of competing techniques on LPBA dataset (mean $\pm$ std, \textbf{$n=9$}). The notation \textbf{Bold} indicates the top result, whereas \underline{underline} denotes the runner-up. The asterisk (*) denotes a statistically significant difference ($p < 0.05$) when comparing our LGANet++ against the respective comparison method using the Wilcoxon signed-rank test. } 
    \label{tab: cross-patient lpba}
    \resizebox{\linewidth}{!}{
    \begin{tabular}{c|ccccc}
        \hline \specialrule{0em}{0.2pt}{0.2pt} \hline
        Methods &DSC(\%) $\uparrow$ &HD95 $\downarrow$  & NJD(\%) $\downarrow$  &Recall (\%) $\uparrow$  & Precision (\%) $\uparrow$ \\ 
        \cline{1-6}
        SyN               &70.03$\pm$0.91* &5.38$\pm$0.23*   &\textbf{0.00007}  &69.26$\pm$0.85* &72.79$\pm$1.18* \\
        VoxelMorph        &66.14$\pm$2.55* &6.04$\pm$0.53*   &0.54  &65.13$\pm$2.64* &69.35$\pm$2.43* \\
        PIViT             &70.58$\pm$1.15* &5.35$\pm$0.30*   &0.15  &70.00$\pm$0.92* &74.08$\pm$0.95* \\
        PRNet++           &69.24$\pm$1.99* &5.64$\pm$0.39*   &0.50  &68.47$\pm$1.99* &72.15$\pm$1.82* \\
        PCNet             &71.85$\pm$1.46* &5.36$\pm$0.35*   &0.39  &71.07$\pm$1.45* &74.70$\pm$1.37* \\
        ModeT             &71.76$\pm$1.37* &5.29$\pm$0.35*   &0.42  &70.53$\pm$1.34* &74.97$\pm$1.10* \\
        CorrMLP           &69.46$\pm$2.29* &5.73$\pm$0.48*   &0.48  &68.56$\pm$2.31* &72.76$\pm$2.02* \\
        GroupMorph        &72.48$\pm$1.04* &5.25$\pm$0.31*   &0.09  &71.31$\pm$0.91* &75.69$\pm$0.86* \\
        RDP               &72.87$\pm$1.18* &5.27$\pm$0.32*   &0.04  &71.79$\pm$1.09 &75.94$\pm$1.10 \\
        \hline
        LGANet (Ours)     &\underline{73.31$\pm$1.02} &\underline{5.13$\pm$0.29}  &0.42  &\underline{71.83$\pm$0.97} &\underline{76.38$\pm$0.94} \\
        LGANet++ (Ours)   &\textbf{73.52$\pm$0.09} &\textbf{5.10$\pm$0.28}   &\underline{0.01} &\textbf{72.24$\pm$0.89} &\textbf{76.80$\pm$0.85} \\
        \hline \specialrule{0em}{0.2pt}{0.2pt} \hline
    \end{tabular}
    }
\end{table}

\begin{table}[h]
    \centering
    \caption{Quantitative results of our approach contrasted with those of competing techniques on IXI dataset (mean $\pm$ std, \textbf{$n=114$}). The notation \textbf{Bold} indicates the top result, whereas \underline{underline} denotes the runner-up. The asterisk (*) denotes a statistically significant difference ($p < 0.05$) when comparing our LGANet++ against the respective comparison method using the Wilcoxon signed-rank test.}
    \label{tab: cross-patient ixi}
    \resizebox{\linewidth}{!}{
    \begin{tabular}{c|ccccc}
        \hline \specialrule{0em}{0.2pt}{0.2pt} \hline

        Methods &DSC(\%) $\uparrow$ &HD95 $\downarrow$ & NJD(\%) $\downarrow$  & Recall (\%) $\uparrow$  & Precision (\%) $\uparrow$  \\ 
        \cline{1-6}
        SyN               &80.51$\pm$2.64*  &4.04$\pm$0.65*  &\textbf{0.0004} &79.33$\pm$2.63* &82.54$\pm$2.19* \\
        VoxelMorph        &79.61$\pm$2.63*  &2.24$\pm$0.20*  &0.52            &78.98$\pm$1.85* &82.70$\pm$2.09* \\
        PIViT             &79.65$\pm$2.07*  &2.28$\pm$0.14*  &0.08            &78.69$\pm$1.62* &81.31$\pm$1.31* \\
        PRNet++           &82.32$\pm$2.06*  &2.15$\pm$0.13*  &0.15            &81.21$\pm$1.53* &84.07$\pm$1.11* \\
        PCNet             &82.65$\pm$1.56*  &2.14$\pm$0.12*  &0.05            &80.47$\pm$1.53* &83.93$\pm$1.37* \\
        ModeT             &82.06$\pm$1.23*  &2.18$\pm$0.13*  &0.15            &81.36$\pm$1.47  &83.29$\pm$1.14* \\
        CorrMLP           &81.43$\pm$2.04*  &2.22$\pm$0.23*  &0.45            &80.20$\pm$1.87*  &83.35$\pm$1.95* \\
        GroupMorph        &\underline{82.84$\pm$1.17}*	 &2.15$\pm$0.10*  &0.11 &81.82$\pm$1.40 &84.04$\pm$0.88* \\
        RDP               &82.78$\pm$1.28*	 &\underline{2.13$\pm$0.10}   &\underline{0.03} &80.56$\pm$1.26* &\underline{84.35$\pm$0.76} \\
        \hline
        LGANet (Ours)     &82.64$\pm$1.32* &2.15$\pm$0.14  &0.05                  &81.46$\pm$1.40 &83.82$\pm$0.89* \\
        LGANet++ (Ours)   &\textbf{83.60$\pm$1.00} &\textbf{2.12$\pm$0.10}  &0.04 &\textbf{82.48$\pm$1.37} &\textbf{85.18$\pm$0.77} \\
        \hline \specialrule{0em}{0.2pt}{0.2pt} \hline
    \end{tabular}
    }
\end{table}

\subsubsection{Cross-patient Registration}
We evaluated our proposed technique for cross-patient registration using two brain datasets.
As quantitatively demonstrated in Table~\ref{tab: cross-patient lpba}, the proposed LGANet++ achieves state-of-the-art performance on the LPBA dataset across multiple evaluation metrics. Specifically, LGANet++ achieves the highest DSC of 73.52\% and the best HD95 of 5.10 mm, outperforming the second-best method RDP by 0.65\% in DSC and demonstrating a 3.23\% improvement in HD95 (5.10 mm vs. RDP's 5.27 mm). Statistical analysis further confirms the significance of these results, with the Wilcoxon Signed-Rank test yielding p-values of 3.9e-3 for DSC and 7.5e-4 for HD95 when compared to RDP. Our method also achieves superior performance in both recall (72.24\%) and precision (76.80\%), significantly different from most comparison methods ($p$-value $<$0.05), indicating its balanced capability in correctly identifying aligned structures while minimizing false positive alignments. 
On the IXI dataset, LGANet++ demonstrates a consistent performance trend with the LPBA dataset, achieving the best overall registration accuracy. As shown in Table~\ref{tab: cross-patient ixi}, it obtains the highest DSC of 83.60\%, significantly outperforming the runner-up method GroupMorph (82.84\%) with a $p$-value of 0.011. While LGANet++ also achieves the best HD95 value of 2.12 mm, the difference from RDP (2.13 mm) is not statistically significant ($p$-value = 0.32), indicating that further improvement is warranted, particularly in challenging regions with ambiguous tissue boundaries. This consistent superiority across datasets further validates the robustness and generalizability of our method in cross-patient brain MRI registration tasks.

\begin{figure*}[tp]
    \centering
    \includegraphics[width=\textwidth]{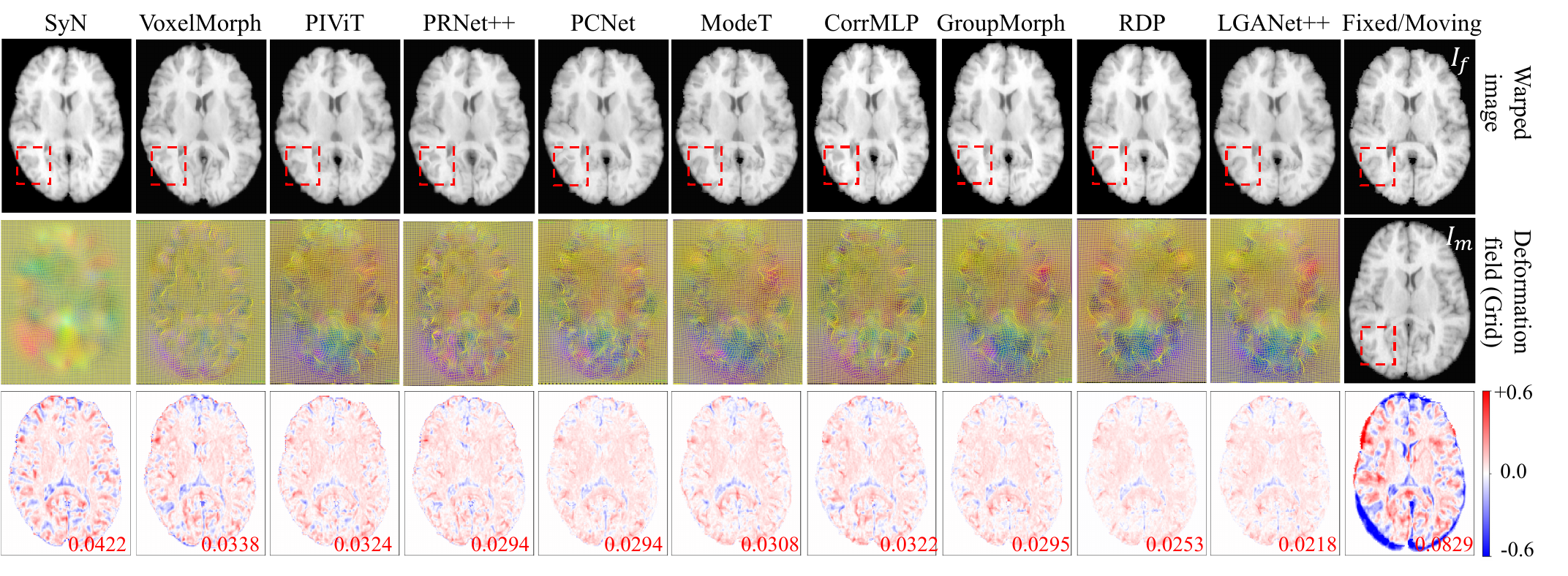}
    \caption{Comparing the visual results obtained by different methods on the LPBA dataset. The third row is the residual maps between warped moving images and fixed images ($I_{residual}=I_f-I_w$), with the mean absolute error ($error=mean(|I_{residual}|)$) placed in the bottom right corner. A cleaner error map indicates a better registration result.}
    \label{fig: visual lpba}
\end{figure*}
\begin{figure}[!t]
    \centering
    \includegraphics[width=0.8\linewidth]{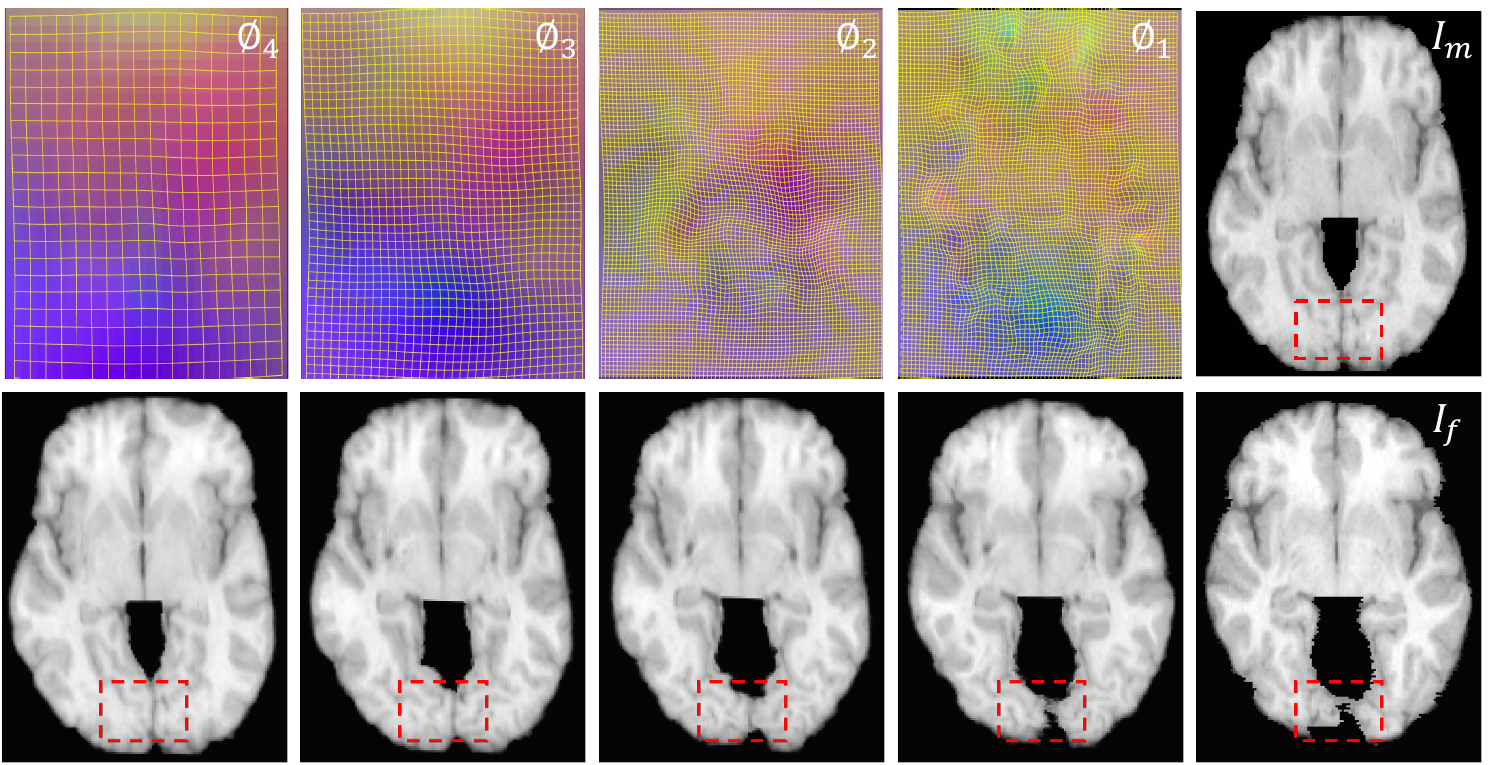}
    \caption{Our coarse-to-fine deformation methodology is demonstrated on the LPBA dataset. The top row displays the intermediate deformation sub-fields ${\phi_{4}, \phi_{3}, \phi_{2}, \phi_{1}}$ predicted at four different resolution levels within the LGANet++ framework (see Fig. 1). Each sub-field is upsampled to the full resolution for visualization. The bottom row shows the corresponding warped moving image at each stage, using a representative 2D slice. The top-right and bottom-right panels present the original moving image and fixed image, respectively, for reference.}
    \label{fig: coarse2fine}
\end{figure}

We also report the percentage of negative Jacobian determinants (NJD) to assess the smoothness and topology-preserving properties of the deformation fields. LGANet++ achieves the lowest NJD on the LPBA dataset($<$0.01\%) and the second-best on the IXI dataset($<$0.04\%) among deep learning methods, indicating that our predictions are not only accurate but also anatomically reliable.
The traditional method SyN demonstrates excellent performance in deformation smoothness (NJD $<$ 0.0004\%), yet its registration accuracy remains suboptimal compared to most of the deep learning-based approaches. More critically, SyN suffers from significant computational inefficiency - while deep learning models complete registration in under 1 second after training, SyN requires approximately 40 seconds to process a single image pair from the LPBA dataset, representing a 40-fold increase in processing time that substantially limits its practical utility in clinical applications requiring rapid results.

\begin{table}[h]
  \caption{Results of external test on the OASIS dataset (mean$\pm$std,\textbf{$n=413$}). The models were trained on the IXI dataset. The notation \textbf{Bold} indicates the top result, whereas \underline{underline} denotes the runner-up. The asterisk (*) denotes a statistically significant difference ($p < 0.05$) when comparing our LGANet++ against the respective comparison method using the Wilcoxon signed-rank test.} 
  \centering
  \label{tab: cross test}
  \resizebox{\linewidth}{!}{
  \begin{tabular}{c|ccccc}
    \hline \specialrule{0em}{0.2pt}{0.2pt} \hline
    Methods&DSC(\%) $\uparrow$ & HD95 $\downarrow$ &NJD(\%)$\downarrow$ &Recal (\%) $\uparrow$ & Precision (\%) $\uparrow$  \\
    \hline
    VoxelMorph      & 70.22$\pm$2.21*   & 3.24$\pm$0.39*  &0.87  &70.02$\pm$2.36* &72.48$\pm$2.25* \\
    PIViT           & 65.04$\pm$2.07*   & 3.58$\pm$0.32*  &0.36  &67.41$\pm$2.31* &64.37$\pm$1.95* \\
    PRNet++         & 73.92$\pm$2.02*   & 2.89$\pm$0.36*  &0.36  &73.51$\pm$2.03* &75.49$\pm$2.14* \\ 
    PCNet           & 73.25$\pm$2.02*   & 2.97$\pm$0.33*  &0.17  &74.36$\pm$2.29* &73.19$\pm$2.39* \\
    ModeT           & 71.37$\pm$2.43*   & 3.18$\pm$0.39*  &0.37  &73.18$\pm$2.39* &72.15$\pm$2.69* \\ 
    CorrMLP         & 73.74$\pm$2.18*   & 2.83$\pm$0.32*  &0.53  &73.86$\pm$2.25* &75.15$\pm$2.20* \\
    GroupMorph      & 71.75$\pm$2.10*   & 3.05$\pm$0.38*  &0.41  &70.75$\pm$2.12* &73.64$\pm$2.24*\\
    RDP             & \underline{75.65$\pm$1.77}*   & \underline{2.79$\pm$0.29}*  &\textbf{0.09} &\underline{74.72$\pm$2.00}* &\underline{76.81$\pm$1.54}* \\
    \hline
    LGANet (Ours)          &75.33$\pm$1.65*     & 2.86$\pm$0.34* &\underline{0.10} &74.42$\pm$2.11* &76.12$\pm$1.87* \\
    LGANet++ (Ours)        & \textbf{76.70$\pm$1.54}   & \textbf{2.71$\pm$0.26}  &\textbf{0.09} &\textbf{76.01$\pm$1.78} &\textbf{78.15$\pm$1.54} \\
    \hline \specialrule{0em}{0.2pt}{0.2pt} \hline
  \end{tabular}
  }
\end{table}

Fig. \ref{fig: visual lpba} provides a visual comparison of different methods on an exemplary slice from the LPBA dataset. LGANet++ yielded more precise warped images and consistently captured internal structures effectively, with finer details preserved in regions highlighted by the red boxes. The difference maps further confirm that our method yields the smallest registration error. The results of VoxelMorph, PIViT, PRNet++, and CorrMLP exhibit a greater divergence from the fixed image. While improvements are observed with ModeT, GroupMorph, and RDP, both methods fell short compared to the effectiveness of LGANet++.

To further illustrate the working mechanism of our approach, Fig. \ref{fig: coarse2fine} demonstrates our coarse-to-fine registration strategy (as described in Sections 2.5 and Fig. \ref{fig: field-refine}) by visualizing the sequence of predicted deformation sub-fields ${\phi_{4}, \phi_{3}, \phi_{2}, \phi_{1}}$. Here, each $\phi_i$ corresponds to an intermediate deformation field estimated by LGANet++ at a specific resolution level (see Fig. \ref{fig: network} for reference). Each $\phi_i$ for $i \in [2,3,4]$ was upsampled to the spatial dimensions of $\phi_1$ and applied independently to warp the moving image. It can be observed that with each successive sub-field, more detailed deformation structures are captured. As a result, through this hierarchical refinement process, the moving image is progressively and accurately aligned with the fixed image.

\begin{figure}[tp]
    \centering
    \includegraphics[width=0.95\textwidth]{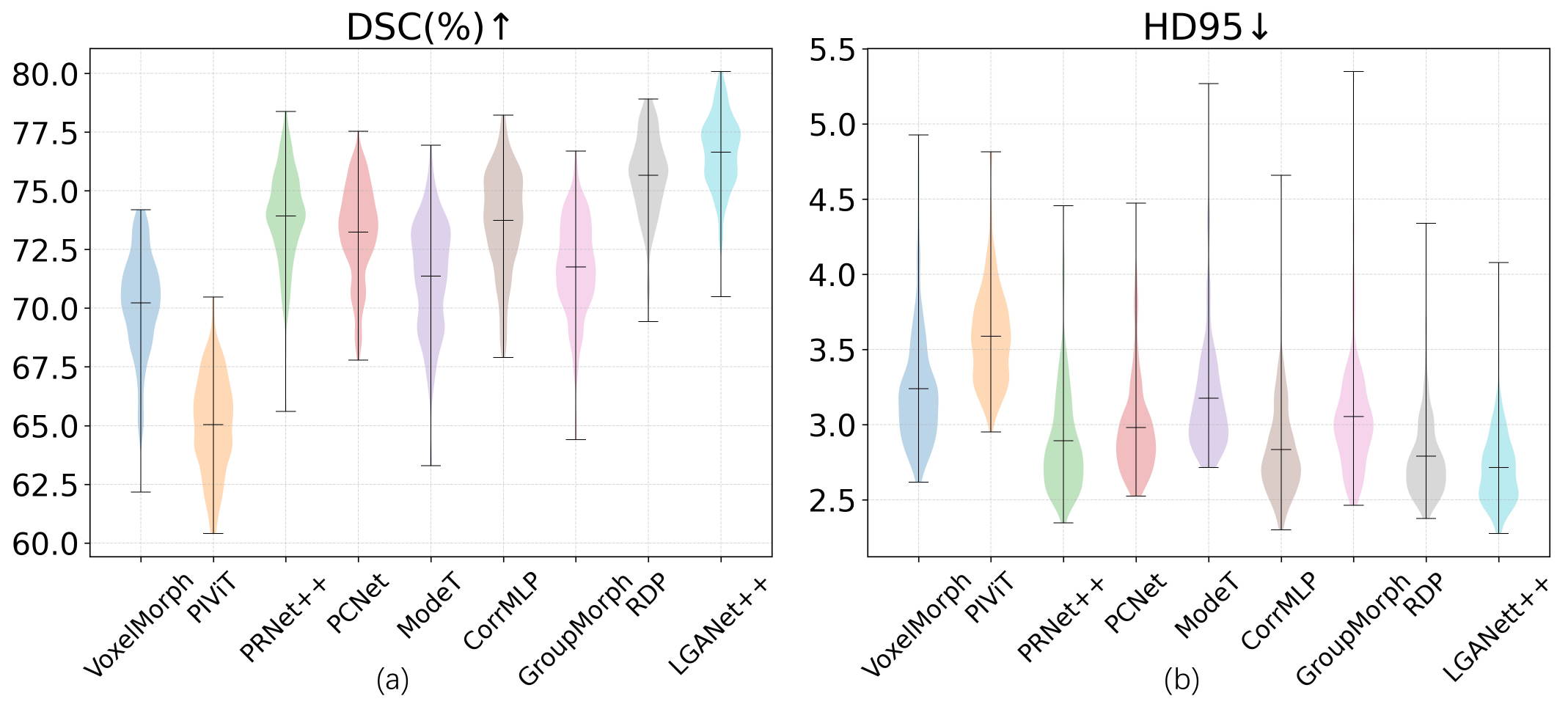}
    \caption{The results of the external OASIS dataset. All models were trained on the IXI dataset but were tested on the OASIS dataset.}
    \label{fig: visual oasis}
\end{figure}

Moreover, to evaluate the generalizability of our approach, we conducted an external validation experiment where models were trained on the IXI dataset and tested on the OASIS dataset. As shown in Table \ref{tab: cross test}, LGANet++ achieved a DSC of 76.70\%, representing a relative improvement of 1.39\% over the second-best method (RDP) on the OASIS dataset. It also attained an HD95 of 2.71, demonstrating strong robustness. All metrics show significant differences, and the $p$-values are all less than 0.05. In comparison, competing methods exhibited notable performance degradation under this domain shift. For instance, GroupMorph—the second-best method on IXI dataset—experiences an 11.09\% decrease in DSC (from 82.84\% to 71.75\%), while RDP declines by 7.13\% (from 82.78\% to 75.65\%). A consistent drop in HD95 further confirms the limited adaptability of these methods to unseen data distributions. Notably, PIViT suffers the most severe reduction, with a 14.61\% decline in DSC, highlighting its sensitivity to dataset variability. These findings underscore the superior generalization capacity of LGANet++, which maintains stable performance across datasets—a critical attribute for real-world clinical deployment. The visual comparison in Fig.~\ref{fig: visual oasis}, which presents the DSC and HD95 metrics, corroborates these quantitative results, clearly illustrating the consistent superiority of our method in cross-domain registration tasks.

\begin{table}[tp]
  \caption{Results obtained on the cross-modal Abdomen CT-MR dataset (mean$\pm$std, \textbf{$n=6$}). The notation \textbf{Bold} indicates the top result, whereas \underline{underline} denotes the runner-up. The asterisk (*) denotes a statistically significant difference ($p < 0.05$) when comparing our LGANet++ against the respective comparison method using the Wilcoxon signed-rank test.}
  \centering
  \label{tab: abdomen CT-MR}
  \resizebox{\linewidth}{!}{
  \begin{tabular}{c|ccccc}
    \hline \specialrule{0em}{0.2pt}{0.2pt} \hline
    Methods & DSC (\%) $\uparrow$ & HD95 $\downarrow$  & NJD(\%) $\downarrow$ & Recall (\%) $\uparrow$ & Precision (\%) $\uparrow$ \\
    \hline
    Initial         & 38.64 & 17.723  \\
    VoxelMorph      & 44.00$\pm$15.61*   & 16.58$\pm$7.70*    & 1.77             &45.20$\pm$15.18*   &44.88$\pm$17.41* \\
    PIViT           & 57.39$\pm$11.70*   & {13.53}$\pm$4.67*  &\textbf{0.58}    &63.22$\pm$13.90*  &54.57$\pm$12.67* \\
    PRNet++         & 43.95$\pm$14.96*   & 16.17$\pm$7.16*    &\underline{1.24}  &45.53$\pm$15.02*  &43.58$\pm$16.87* \\
    PCNet           & 62.60$\pm$14.54*   & 13.13$\pm$6.31*    &1.42              &61.83$\pm$17.88*  &67.12$\pm$11.06* \\
    ModeT           & 72.68$\pm$8.37*    & 10.96$\pm$3.13*    &1.72              &78.90$\pm$11.20*  &70.43$\pm$9.88* \\
    CorrMLP         & 55.80$\pm$16.13*   & 14.71$\pm$5.91*    &1.54              &59.00$\pm$18.02*  &54.94$\pm$14.70* \\
    GroupMorph      & 66.01$\pm$16.12*   & 13.29$\pm$7.16*    &2.11              &68.36$\pm$14.44*  &63.74$\pm$12.04* \\
    RDP             & 75.65$\pm$10.73*   & \underline{8.39$\pm$3.34}*     &1.83  &79.08$\pm$10.49*  &71.91$\pm$10.79* \\
    \hline
    LGANet (Ours)          &\underline{77.36$\pm$10.59}*    & 9.06$\pm$3.35*    &1.37  &\underline{80.42$\pm$8.77}*  &\underline{73.80$\pm$11.32} \\
    LGANet++ (Ours)           & \textbf{80.28}$\pm$7.86   & \textbf{6.43$\pm$2.44}   &1.65  &\textbf{84.72$\pm$9.38}  &\textbf{74.62$\pm$9.12} \\
    \hline \specialrule{0em}{0.2pt}{0.2pt} \hline
  \end{tabular}
  }
\end{table}

\begin{figure}[!t]
    \centering
    \includegraphics[width=0.95\linewidth]{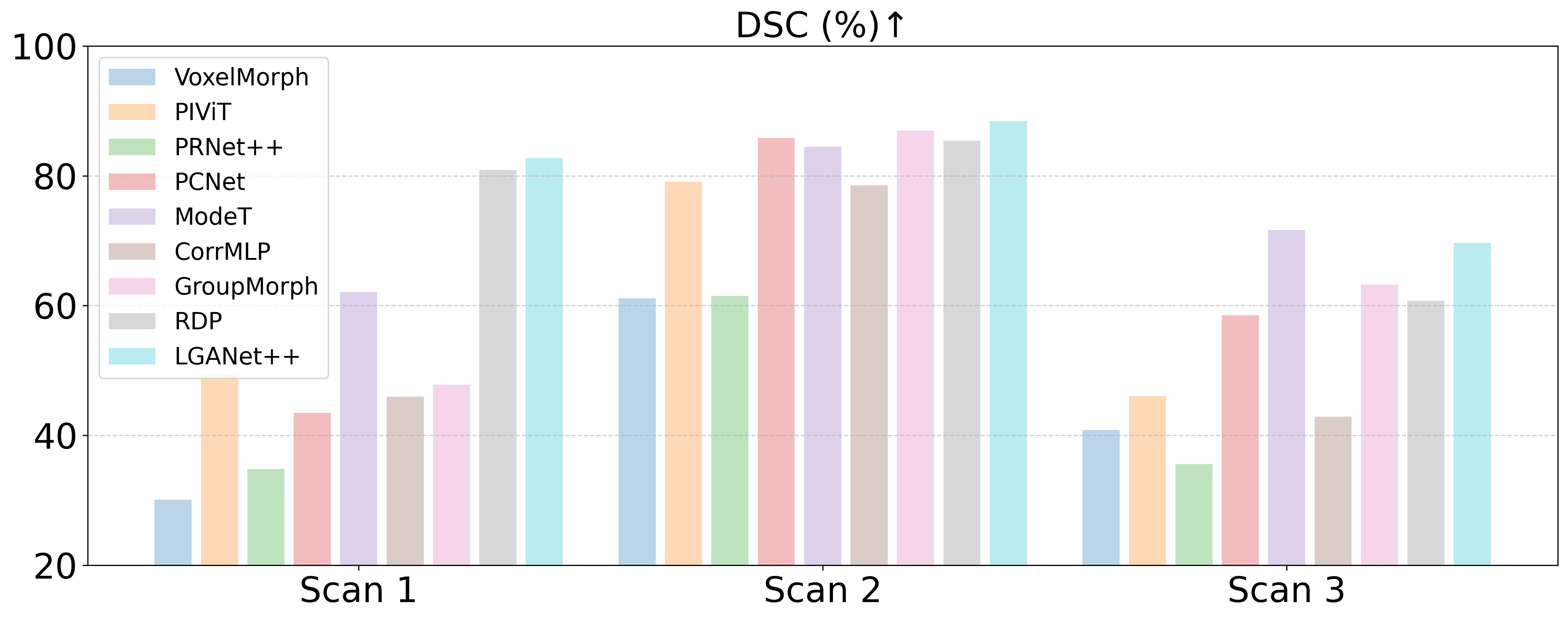}
    \caption{Visualization results of various registration techniques on three representative scans from the Abdomen CT-MR dataset.}
    \label{fig: visual ct-mr}
\end{figure}

\subsubsection{Cross-modal Registration}
We evaluated the performance of our method on the challenging Abdomen CT-MR cross-modal registration task using the corresponding dataset. As depicted in Table \ref{tab: abdomen CT-MR}, LGANet++ demonstrated superior performance to all other methods, notably achieving a DSC of 80.28\%, which represents a relative improvement of 6.12\% over the runner-up, RDP. Furthermore, LGANet++ shows significant improvement in boundary alignment accuracy, attaining an HD95 value of 6.43 mm. In contrast, both VoxelMorph and PRNet++ underperformed, with a DSC of 0.44. Despite LGANet++ achieving better scores in DSC and HD95 metrics, PIViT exhibits the smallest NJD score, with a value of 0.58. LGANet++ also achieves the highest recall (84.72\%) and precision (74.62\%) among all compared methods, demonstrating its balanced capability in both correctly identifying aligned anatomical structures and minimizing false positive alignments. This represents a notable improvement over the runner-up method, RDP, which attains 79.08\% recall and 71.91\% precision in this challenging cross-modal registration task.

As illustrated in Fig. \ref{fig: visual ct-mr}, a comprehensive comparison of registration results across three representative test scans validates the superior accuracy and robustness of LGANet++ on the cross-modal registration scenarios.

\begin{table}[h]
  \centering
  \caption{Result on the cross-time Lung CT dataset (mean$\pm$std, \textbf{$n=10$}). The notation \textbf{Bold} indicates the top result, whereas \underline{underline} denotes the runner-up. The asterisk (*) denotes a statistically significant difference ($p < 0.05$) when comparing our LGANet++ against the respective comparison method using the Wilcoxon signed-rank test.}
  \label{tab: lung CT}
  \resizebox{\linewidth}{!}{
  \begin{tabular}{c|cccccc}
    \hline \specialrule{0em}{0.2pt}{0.2pt} \hline
    Methods & DSC(\%) $\uparrow$ & HD95 $\downarrow$ &TRE $\downarrow$ & NJD(\%) $\downarrow$ & Recall (\%) $\uparrow$ & Precision (\%) $\uparrow$ \\
    \hline

    VoxelMorph      & 92.71$\pm$4.73*  &15.07$\pm$6.98*  &10.36$\pm$4.02*  & 0.42 &88.53$\pm$8.57* &97.85$\pm$0.60 \\
    PIViT           & 96.51$\pm$2.00*  &5.54$\pm$5.23*   &3.09$\pm$1.63*   & 0.05 &95.67$\pm$3.28* &97.53$\pm$1.29 \\
    PRNet++         & 94.36$\pm$4.08*  &11.37$\pm$10.25* &6.25$\pm$3.67*   & 0.59 &91.03$\pm$7.57* &\textbf{98.37$\pm$0.55} \\
    PCNet           & 96.16$\pm$2.44*  &6.80$\pm$5.69*   &2.67$\pm$1.82*   & 0.10 &95.37$\pm$3.76* &97.26$\pm$1.65 \\
    ModeT           & 96.68$\pm$1.69*  &6.05$\pm$4.51*   &2.48$\pm$1.03*   & 0.11 &\underline{96.18$\pm$2.52}* &97.26$\pm$1.83 \\
    CorrMLP         & 95.36$\pm$2.42*  &7.38$\pm$5.60*   &3.57$\pm$2.43*   & 0.13 &94.71$\pm$3.94* &96.59$\pm$2.25* \\
    GroupMorph      & 96.33$\pm$2.16*  &6.75$\pm$5.66*   &2.59$\pm$1.68*   & 0.60 &95.42$\pm$2.70* &97.44$\pm$2.06 \\
    RDP             & \underline{96.92$\pm$1.51}  &\textbf{4.49$\pm$3.27}  &\underline{2.32$\pm$0.77}*  & \underline{0.01} &95.71$\pm$1.33* &97.43$\pm$1.38 \\
    \hline
    LGANet (Ours)         &96.74$\pm$1.59*    &5.12$\pm$3.97*    &2.45$\pm$1.27*    &0.06 &96.06$\pm$2.17* &97.15$\pm$1.44 \\
    LGANet++ (Ours)           & \textbf{97.61$\pm$1.21}  &\underline{4.50$\pm$3.84}  &\textbf{2.02$\pm$0.64}  & \textbf{0.002} &\textbf{97.20$\pm$2.03} &\underline{97.86$\pm$1.56} \\
    \hline \specialrule{0em}{0.2pt}{0.2pt} \hline
  \end{tabular}
  }
\end{table}

\subsubsection{Cross-time Registration}
We carried out a study on the Lung CT dataset to assess the effectiveness of our proposed technique on the cross-time registration scenario. The corresponding results are presented in Table \ref{tab: lung CT}. Our method achieves the highest DSC (97.71\%) and the lowest TRE (2.02 mm) and NJD (0.002\%) scores, indicating the most consistent registration performance. Notably, our method achieves a 0.71\% improvement in DSC and a 12.9\% enhancement in TRE compared to the second-best approach, RDP, demonstrating superior performance in both overlap-based and landmark-based registration accuracy.
In terms of Recall and Precision, LGANet++ attains the highest Recall score of 97.20\%, while achieving competitive Precision of 97.86\%, second only to PRNet++ (98.37\%). This indicates LGANet++ excels at correctly identifying aligned regions while maintaining high registration accuracy, and our method is significantly superior to the comparison methods in the Recall metric, with $p$-values all less than 0.05. While PRNet++ attained the highest Precision, its significantly lower Recall (91.0\% vs. our 97.20\%) indicates a substantial number of false negative alignments, particularly in regions with large deformations where its registration capability appears limited. This contrast highlights LGANet++'s superior ability to maintain comprehensive anatomical coverage while ensuring accurate alignment, demonstrating balanced performance across both sensitivity and precision metrics.
Although our method improves the registration results, the improvement in DSC is not a significant difference compared to the second-place RDP ($p$-value of 0.28). While RDP achieves the best performance in the HD95 metric, it's only 0.01 better than LGANet++, and there is no significant difference between the two ($p=0.46$). It is noteworthy that LGANet++ yields a substantially lower NJD value (0.002\% vs. RDP's 0.01\%). 
This improvement in deformation field quality indicates that our method produces more reliable and physiologically plausible transformations, with better preservation of topological structures and reduced folding artifacts, despite the comparable performance in overlap-based metrics.

\begin{figure}[t]
    \centering
    \includegraphics[width=\linewidth]{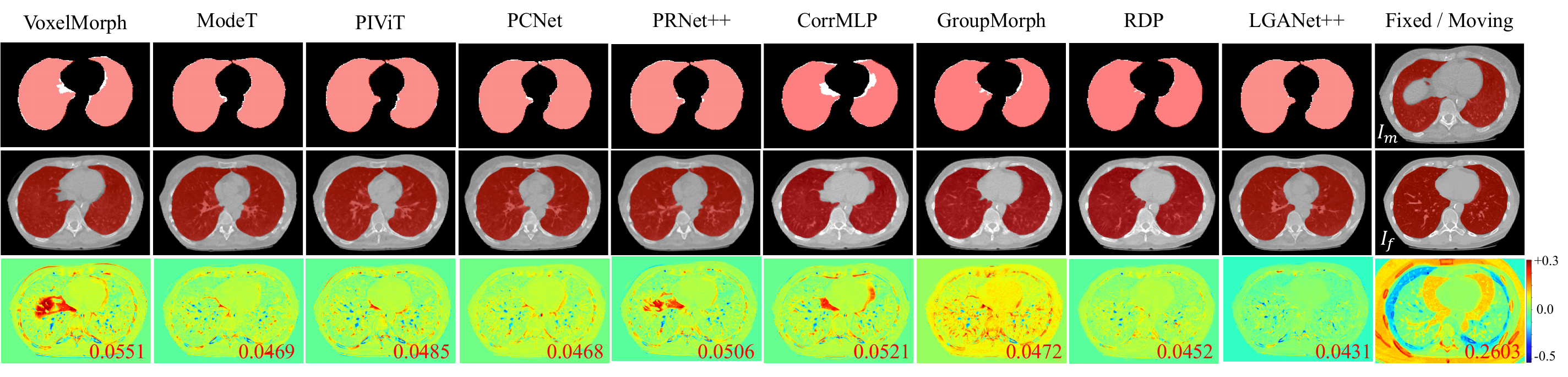}
    \caption{Visualization results of various registration techniques applied to the Lung CT dataset. The final column presents the moving and fixed images, along with the residual images calculated as ($I_{residual} = I_f - I_m$). In the remaining columns, the first row displays the results of the different registration techniques overlaid on the fixed labels, with white denoting the fixed label and pink indicating the registration result. The second row illustrates the registration results of the different methods superimposed on the warped images. Meanwhile, the third row reveals the respective residual images between the fixed and warped images, expressed as ($I_{residual} = I_f - I_w$). The red numbers placed in the bottom right corner are mean absolute errors.}
    \label{fig: visual-lung CT}
\end{figure}

Fig. \ref{fig: visual-lung CT} presents visual examples of the comparative analysis. It is noteworthy that our warped images exhibited the greatest similarity to the fixed image, resulting in cleaner residuals. 

In summary, the comprehensive experiments conducted on five publicly available datasets across three distinct registration scenarios—cross-patient, cross-time, and cross-modal—demonstrate the consistent superiority of our approach over state-of-the-art methods. Our approach not only achieves significant improvements in registration accuracy (as reflected by DSC, HD95, TRE, and other metrics) but also exhibits strong generalization and robustness under domain shifts. These results validate the effectiveness of the proposed local-global attention mechanism, multi-scale fusion strategy, and hierarchical refinement process in handling complex and varied deformation patterns, underscoring the potential of LGANet++ for real-world clinical applications.

\section{Discussion}\label{discussion}
In this study, we have developed LGANet++, a local-global attention network that advances unsupervised deformable image registration by integrating pyramid registration, local-global attention mechanisms, and multi-scale feature fusion into a unified coarse-to-fine framework. The main contributions of this study are threefold: (1) methodologically, we introduced a Local-Global Attention Module (LGAM), a Feature Interaction and Fusion Module (FIFM), and a Multi-Scale Fusion Module (MSFM), which jointly enhance both local correspondence and global contextual coherence during registration; (2) experimentally, LGANet++ consistently achieved state-of-the-art performance across diverse tasks, showing accuracy improvements of 1.39\% in cross-patient registration, 0.71\% in cross-time registration, and 6.12\% in cross-modal CT-MR registration tasks. Particularly in external validation, LGANet++ showed the smallest performance drop, evidencing outstanding robustness to distribution shifts; and (3) clinically, the demonstrated capability to handle cross-patient, cross-time, and cross-modal registration highlights its potential utility in applications such as intraoperative navigation and longitudinal disease monitoring. 
It is worth noting that methods lacking an attention mechanism, such as VoxelMorph and PRNet++, consistently underperformed across the LPBA, IXI, Lung CT, and Abdomen CT-MR datasets, particularly in cross-modal registration tasks. These observations underscore the importance of incorporating attention mechanisms for achieving robust and accurate image registration.
Despite these promising findings, further efforts are needed to enhance clinical applicability, mitigate existing limitations, and explore future work for improvement.

\subsection{Model Performance}
Ablation analysis (Table 1) reveals that while the total parameters increase from 2.936M (Model 1) to 24.353M (Model 4), the performance gains are primarily driven by architectural innovation rather than raw capacity expansion. Notably, the integration of MSFM in Model 4 introduces only a 14\% parameter increment relative to Model 3, yet yields a substantial 0.95\% boost in DSC—a level of efficiency unattainable through increase of the parameters.
Exist  literature indicates that simply increasing network depth or filter counts often results in diminishing returns, training instability, and overfitting \citep{he2016deep}. In contrast, a meticulously designed structure can more effectively capture complex latent data distributions. Furthermore, the consistently superior performance of LGANet++ across diverse datasets (e.g., IXI, OASIS, Abdomen CT-MR, and Lung CT) underscores the stablity of its structural design in handling heterogeneous anatomical variations.

\subsection{Generalizability and Robustness}
The consistent state-of-the-art performance of LGANet++ across five publicly available datasets with vastly different characteristics underscores its strong generalization capability. Our method effectively handled the nuanced challenges inherent in each scenario: the detailed anatomical variability in cross-patient brain registration (LPBA and IXI datasets), the significant non-rigid motion in lung CT scans, and the pronounced intensity and contrast discrepancies in cross-modal Abdomen CT-MR tasks. This robustness can be attributed to the synergistic design of its core components. The local-global attention mechanism (LGAM) captures both fine-grained local correspondences and long-range contextual relationships, which is crucial for aligning structures that vary significantly in appearance across modalities or patients. Furthermore, the multi-scale fusion and coarse-to-fine optimization strategy ensure that the registration process is guided by robust high-level semantic information before being refined with precise low-level details, making the framework adaptable to a wide spectrum of deformation magnitudes and image qualities. The external cross-dataset validation, where models trained on IXI were tested on OASIS, provides compelling evidence for this generalizability. LGANet++ exhibited the smallest performance drop in this challenging setting, indicating that it learns meaningful and transferable representations of anatomical alignment rather than merely overfitting to the intensity distribution of a specific dataset.

\subsection{Clinical Implications}

Our method demonstrates strong performance across three various registration scenarios—cross-patient, cross-patient, and cross-modal—highlighting its broad applicability in clinical practice.
In cross-patient registration, the method can facilitate population-based studies and atlas-based segmentation by aligning images from different individuals, enabling the construction of standardized anatomical templates and supporting neuroimaging research on brain development and degenerative diseases \citep{che2025nested}. Additionally, it can enhance surgical planning systems by fusing preoperative images from multiple patients to create comprehensive anatomical references.
For cross-time registration, the approach proves valuable in monitoring disease progression in chronic conditions such as Alzheimer's disease or tumor evolution in oncology \citep{di2024graph}. It enables precise measurement of organ volume changes in longitudinal studies and supports assessment of treatment efficacy in clinical trials. The method's ability to handle large deformations also benefits radiation therapy by accurately tracking anatomical changes during fractionated treatment.
In cross-modal registration, the technique enables comprehensive multi-parametric analysis by fusing complementary information from different imaging modalities \citep{yang2025rethinking}. Specifically, it can integrate functional PET data with anatomical MRI to improve tumor delineation in oncology, combine CT angiography with MR perfusion for stroke evaluation, and fuse ultrasound with MRI for intraoperative guidance. These capabilities significantly enhance diagnostic confidence and treatment planning accuracy.
The unsupervised nature of our method eliminates the dependency on manually annotated data, which are particularly scarce for multi-modal and longitudinal applications, thereby improving its practicality and adoption potential in diverse clinical workflows.

\subsection{Limitations}
Despite its strong performance, the proposed method has certain limitations that warrant acknowledgment. On the IXI and Abdomen CT-MR datasets, our method occasionally produced deformation fields with a higher fraction of negative Jacobian determinants compared to some competing methods, indicating local non-diffeomorphic behavior. Such topological inconsistencies could potentially limit the method's applicability in scenarios requiring strict invertibility, such as in anatomy studies that track longitudinal changes. Additionally, the architectural design—particularly the repeated application of the Feature Interaction and Fusion Module (FIFM) during the decoding phase—results in increased model complexity and substantial GPU memory consumption. This elevated computational cost may present a practical barrier for deployment in resource-constrained clinical environments, as the translation of technical advancements into clinical practice requires not only high accuracy but also computational efficiency, and seamless integration into existing workflows. 

\subsection{Future Work}
Building upon the promising results of this study, several avenues for future work present themselves. To address the current limitation in deformation smoothness, we plan to investigate the integration of biomechanical constraints or adversarial regularization techniques to enforce more physiologically plausible diffeomorphic transformations without compromising registration accuracy. Furthermore, enhancing the model's efficiency is a critical goal; we will explore architectural simplifications, such as model distillation and pruning, to develop a lightweight variant suitable for deployment on standard clinical hardware. Ultimately, the clinical translation of this technology is paramount. Future efforts will focus on rigorous clinical validation in real-world scenarios such as intraoperative navigation, longitudinal treatment monitoring, and multi-modal fusion. Additionally, we aim to develop a more comprehensive analytical tool by integrating the registration framework with segmentation networks in a joint learning paradigm, which could mutually enhance the performance of both tasks and provide a unified solution for automated image analysis pipelines.

\section*{CRediT authorship contribution statement}
\textbf{Zhengyong Huang: }Conceptualization, Data curation, Formal analysis, Investigation, Methodology, Software, Validation, Visualization, Writing – original draft, Writing – review \& editing.  
\textbf{Xingwen Sun: }Investigation, Data curation, Resources.
\textbf{Xuting Chang: }Investigation, Data curation, Resources.
\textbf{Niang Jiang: }Conceptualization, Formal analysis, Visualization.
\textbf{Yao Wang: }Formal analysis, Resources, Funding acquisition.
\textbf{Jianfei Sun: }Resources, Supervision, Funding acquisition.
\textbf{Hongbin Han: }Resources, Supervision, Funding acquisition.
\textbf{Yao Sui: }Conceptualization, Formal analysis, Writing – review \& editing, Project administration, Supervision, Funding acquisition.

\section*{Acknowledgments}
Research reported in this publication was supported in part by Beijing Natural Science Foundation under Award Number L258055; in part by the Major Program of the National Natural Science Foundation of China under Award Numbers 62394310 and 62394312, and in part by the National Natural Science Foundation of China under Award Number 82201601.

\section*{Declaration of competing interest}
The authors declare that they have no known competing financial interests or personal relationships that could have appeared to influence the work reported in this paper.

\section*{Data availability}
The link to our code has been provided in the manuscript. All the used datasets are publicly available.

\bibliographystyle{cas-model2-names}\biboptions{authoryear}

\bibliography{refs}
\end{document}